\documentclass{epl}%
\usepackage{amssymb}
\usepackage{psfig}
\usepackage{amsmath}
\usepackage{amsfonts}
\usepackage{graphicx}%
\begin{document}

\title{Roughness correction to the Casimir force :\\Beyond the Proximity Force Approximation}
\author{Paulo A. Maia Neto \inst{1} 
\and A. Lambrecht\inst{2}
\and S. Reynaud\inst{1} 
\thanks{E-mail: \email{reynaud@spectro.jussieu.fr} ; 
Url: \email{www.spectro.jussieu.fr/Vacuum} } }
\shortauthor{P. A. Maia Neto \etal}
\shorttitle{Roughness correction to the Casimir force}
\institute{ \inst{1} Instituto de F\'{\i}sica, UFRJ, Caixa Postal 68528, 21945-970 Rio de Janeiro RJ, Brazil\\
\inst{2} Laboratoire Kastler Brossel,
CNRS, ENS, UPMC case 74, Campus Jussieu, 75252 Paris, France}

\recff{}{}
\maketitle

\begin{abstract}
We calculate the roughness correction to the Casimir effect in the parallel plates geometry,
for metallic plates described by the plasma model.
The calculation is perturbative in the roughness amplitude, 
with arbitrary values for the plasma wavelength, the plate separation and the 
roughness correlation length.
The correction is found to be always larger than the result obtained in the 
Proximity Force Approximation.
\end{abstract}

High precision measurements of the Casimir force~\cite{Casimir}
have been reported during the last years~\cite{experimental}. 
They allow for an accurate theory/experiment comparison 
\cite{theory_exp} and, consequently, for a search for new weak forces
with submillimetric ranges \cite{search_nwf}.
Theoretical predictions have to take into account the differences between
realistic models of the experiments and the ideal configuration initially 
considered by Casimir. Most experiments are performed with a plane-sphere
geometry rather than a plane-plane geometry. Temperature corrections 
to be added to the vacuum contribution play an important role 
when the distance $L$ between the plates is above 1$\mu$m.
Finite conductivity and roughness of the metallic plates used in the 
experiments provide the major corrections for the distances of the order of 
a few hundred nanometers probed by the most accurate experiments. 
The spatial variations of the surface potential also affect the force 
measurement \cite{Speake03}.

All these effects must be considered simultaneously since they affect a
single observable, the Casimir energy. For example,
the thermal and plasma corrections cannot simply be multiplied in the 
intermediate range where both effects are noticeable \cite{Genet00}
because they are in fact correlated to each other.
In this letter, we study the correlation arising between the conductivity 
and roughness corrections at the short distances where both are appreciable. 
To this aim, we describe the optical response of the metallic plates 
by a plasma model with a dielectric function $\varepsilon=1-\omega_P^2/\omega^2$
where $\omega_P$ is the plasma frequency. 
We evaluate the roughness effect perturbatively in the roughness
amplitude. This effect then depends on the hierarchy between the 
other relevant length scales, namely the plate separation $L$, the plasma 
wavelength $\lambda_P=2\pi c/\omega_P$ and the correlation length $\ell_C$ 
which characterizes the roughness spectrum.

We first consider a plane-plane geometry and define the surface profiles by the functions
$h_i(x,y)$ $(i=1,2)$ giving the local heights with respect to the mean separation $L$ 
along the $z$ direction. These functions are defined so that they have zero averages.
We consider the case of stochastic roughness characterized by spectra 
\begin{equation}
\sigma_{ii}(\mathbf{k}) = \int\mathrm{d}^{2}\mathbf{r}%
\,e^{-i\mathbf{k}\cdot\mathbf{r}}\langle h_i(\mathbf{r})%
\,h_i(\mathbf{0})\rangle,\;\;i=1,2. 
\end{equation}
The surface $A$ of the plates is supposed to contain many correlation areas, allowing 
us to take ensemble or surface averages interchangeably.  
The two plates are considered to be made of the same metal and the crossed 
correlation between their profiles is neglected. 

The variation of the Casimir energy $E_{PP}$ is calculated to second order 
in the perturbations $h_i$, leading to the following expression 
for the roughness correction~\cite{EPL} 
\begin{eqnarray}
\label{main}
&&\delta E _{PP}  =\int\frac{\mathrm{d}^{2}\mathbf{k}}{4\pi^{2}}
G\left(\mathbf{k}\right)  \sigma(\mathbf{k}) \\
&&\sigma(\mathbf{k})=\sigma_{11}(\mathbf{k})+\sigma_{22}(\mathbf{k}).
\nonumber
\end{eqnarray}
With our assumptions, the spectrum $\sigma(\mathbf{k})$ fully characterizes the
roughness of the two plates. The correlation length $\ell_C$ is defined as the 
inverse of its width.
The response function $G\left(\mathbf{k}\right)$ then describes the 
spectral sensitivity to roughness of the Casimir effect. 
Symmetry requires that it only depends on $k=|\mathbf{k}|$.
The dependence of $G$ on $k$ reflects that not only the roughness amplitude 
but also its spectrum plays a role in diffraction on rough surfaces~\cite{roughness}. 
It is only at the limit of smooth surface profiles $k \to 0$
that the effect of roughness may be calculated from the 
Proximity Force Approximation (PFA)~\cite{PFA} by averaging the `local' 
distances over the surface of the plates.

In previous discussions of the roughness corrections~\cite{EPL}, the
sensitivity function $G\left(k\right)$ was analyzed only in the two cases
of short ($L\ll \lambda_P$) and long distances ($L\gg \lambda_P$). 
For the short range limit, it was deduced from earlier 
calculations of Maradudin and Mazur~\cite{Maradudin}.
In the long range limit, it was derived from the evaluation by Emig {\it et al.}~\cite{Emig} 
of the effect of corrugation of a perfectly reflecting plate.  
In the present letter, we give the results of a new evaluation of $G\left(k\right)$ valid
for arbitrary separations $L$.
This evaluation relies on calculations of non-specular reflection coefficients associated 
with rough plates, taking into account the roughness-induced coupling between 
Transverse Electric (TE) and Transverse Magnetic (TM) polarizations. 
The full calculations will be presented in a longer paper~\cite{tbp}. 
Here, we discuss their results which allow us to obtain the roughness correction at any distance, in 
particular at the intermediate distances corresponding to most experiments. 

Before entering this discussion, let us emphasize that all these results may be applied to the 
analysis of the plane-sphere geometry employed in most experiments. To this aim, we use the PFA 
to relate this plane-sphere geometry to the plane-plane configuration, which was taken as 
the benchmark for our perturbative calculation. 
We thus obtain the relative correction of the force $F_{PS}$ in the plane-sphere geometry 
from that evaluated in (\ref{main}) for the plane-plane geometry~\cite{EPL}
\begin{equation}
\label{fps}
\Delta \equiv \frac{\delta F_{PS}}{F_{PS}} = \frac{\delta E_{PP}}{E_{PP}}.
\end{equation}
Note that applying the PFA to the study of the sphere-plane geometry only requires 
the sphere radius $R$ to be sufficiently large. 
Besides the usual requirement $R \gg L$, it is also necessary to assume 
$RL \gg \ell_C^2$, so that many correlation areas are included 
in a given nearly-plane local section of the sphere.
In contrast, using the PFA to calculate the roughness correction requires the 
correlation length to be larger than the separation $\ell_C \gg L$. 
This is clearly a more restrictive condition.
In the present work, we assume the PFA validity conditions to be obeyed 
for plane-sphere geometry but not necessarily for roughness.

According to (\ref{main}), the relative roughness correction (\ref{fps}) 
is obtained by integrating the ratio $G\left(k\right)/E_{PP}$ 
over the roughness spectrum $\sigma\left(\mathbf{k}\right)$. 
This ratio is plotted on Fig.~1 as a function of the roughness wavevector $k$ 
for several different values of the distance $L$.
As for all numerical examples considered below, we take $\lambda_P=136$nm which 
corresponds to gold covered plates. 
As expected, the relative roughness correction is larger for shorter distances. 
In the following, we discuss the values of $G$ in the limit $k \to 0$
corresponding to the PFA. We then come to the main result of this work, that is to
say the $k$-dependence of $G$ which reveals the departure of the effect of
roughness from its PFA description.
\begin{figure}[ptb]
\centerline{\psfig{figure=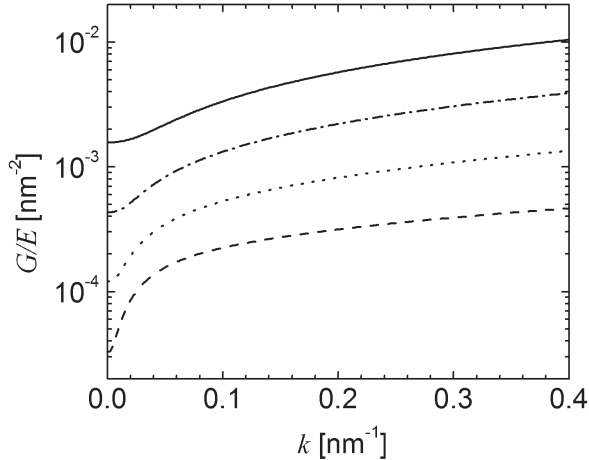,width=9cm}}
\caption{Variation of $G/E_{PP}$
versus $k$ for the distances $L=50 {\rm nm}$ (solid line), 
$L=100 {\rm nm}$ (dashed-dotted line), $L=200 {\rm nm}$ (dotted line), 
and $L=400 {\rm nm}$ (dashed line) for $\lambda_P=136 {\rm nm}.$ }
\label{Rho}
\end{figure}

The PFA result is recovered as a consequence of the following identity
\begin{equation}\label{G0}
G\left(k \to 0\right)=\frac{E_{PP}''(L)}{2}
\end{equation}
where the derivative is taken with respect to the plate separation $L$.
This identity is obeyed by our result for arbitrary values of $L$
and $\lambda_P$ \cite{foot}.
If we now suppose that the roughness spectrum $\sigma(\mathbf{k})$ is 
included inside the PFA sector where $G\left(k\right) \simeq G\left(0\right)$, 
$G$ may be replaced by its limiting value (\ref{G0}) and factored out of the 
integral (\ref{main}) thus leading to the PFA expression~\cite{EPL} 
\begin{eqnarray}\label{eqPFA}
&&\Delta = \frac{E''_{PP}(L)}{2 E_{PP}} a^2 \\
&&a^2 = \int\frac{\mathrm{d}^{2}\mathbf{k}}{4\pi^{2}} \sigma(\mathbf{k}) 
\equiv \langle h_1^2 + h_2^2 \rangle. \nonumber
\end{eqnarray}
In this PFA limit, the correction depends only on the variance $a^2$ of the roughness 
profiles, that is also the integral of the roughness spectrum. 

In the general case in contrast, the sensitivity to roughness depends on the 
wavevector $k$. This key point is emphasized by introducing a new function $\rho$ 
which measures the deviation from the PFA~\cite{EPL}
\begin{equation}
\rho(k) = \frac{G(k)}{G(0)}.
\end{equation}
This function is plotted on Fig.~2 as a function of $k$ for several values of $L$.
It is almost everywhere larger than unity, which means that the PFA systematically
underestimates the roughness correction.
The inlet shows $\rho$ for small values of $k$ where the PFA is a good approximation 
for the shortest distances, for example $L=50$nm.
To give a number illustrating the deviation from the PFA, we may say 
that $\rho\simeq 1.6$ for $L=200$nm and $k = 0.02 {\rm nm}^{-1}$,
which means that the exact correction is $60\%$ larger than the PFA result
for this intermediate separation and a typical roughness wavelength 
$2\pi/k \simeq 300$nm.

\begin{figure}[ptb]
\centerline{\psfig{figure=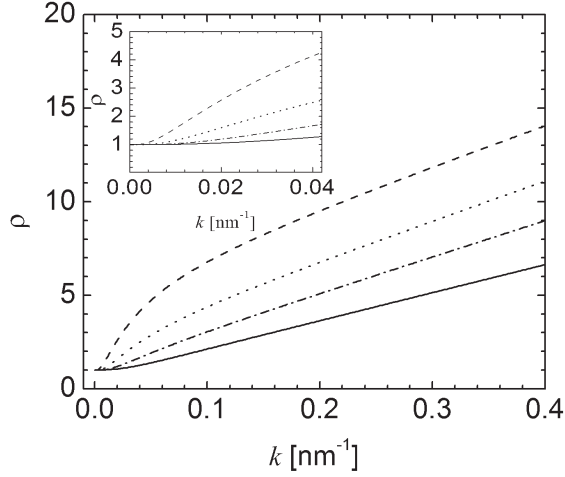,width=9cm}}\caption{Variation of $\rho$
versus $k$ for several values of $L$ (same conventions as on Fig.~1).}
\end{figure}

Fig.~2 indicates that $\rho(k)$ grows linearly for large values of $k$. 
This is in fact a general prediction of our full calculations
for arbitrary values of $L$ and $\lambda_P$: 
\begin{equation}
\rho(k)= \alpha\, k \quad {\rm for}\quad k \gg \omega_P/c, 1/L.
\end{equation}
The dimensionless parameter $\alpha/L$ depends on
$K_P=\omega_P L/c=2\pi L/\lambda_P$ only, and is given by  
\begin{align}\label{high-k}
\alpha=&\frac{\hbar A}{(2\pi)^2 L^4 G(0)}
\int_0^{\infty} dK K \int_0^K d\Omega
\frac{K_P^2}{2\Omega^2+K_P^2}\\
\times & \Biggl\{K f_{\rm TE}+\frac{2(K^2-\Omega^2)^2-K_t^2(2K^2-3\Omega^2)}
 {(K K_t)^2-(K^2-\Omega^2)^2}
K f_{\rm TM}\Biggr\}.
\nonumber
\end{align}
The dimensionless integration variable 
$K$ represents the $z$ component of the imaginary wavevector 
multiplied by $L$ \cite{Lambrecht2000}, and $K_t=\sqrt{K^2+K_P^2}$ corresponds to the $z$ 
component of the imaginary wavevector inside the metallic medium. 
We denote similarly $\Omega$ the imaginary frequency multiplied by $L/c$.
$f_{\epsilon}$ are the loop functions describing the optical response
of the cavity for the two orthogonal polarizations $\epsilon= {\rm TE}, {\rm TM}$: 
\begin{equation}
f_{\epsilon}=\frac{r_{\epsilon}^2 \exp(-2K)}{1-r_{\epsilon}^2 \exp(-2K)}.
\end{equation}
The corresponding reflection coefficients are given by  
\begin{eqnarray}
&&r_{\rm TE} = -\frac{K_t-K}{K_t+K} \\
&&r_{\rm TM} = \frac{\left(1+\frac{K_P^2}{\Omega^2}\right)K-K_t}
{\left(1+\frac{K_P^2}{\Omega^2}\right)K+K_t}. \nonumber
\end{eqnarray}

In Fig.~3, we plot the coefficient $\alpha$ as a function of $L$,
still with the plasma wavelength of gold $\lambda_P=136$nm.
At the limit of short distances, we recover from (\ref{high-k}) 
our previous result \cite{EPL} 
\begin{equation} \label{kgde-plasmon}
\alpha = 0.4492 L \quad {\rm for} \quad k^{-1}\ll L\ll \lambda_P.
\end{equation}
At the limit of large distances, the angular coefficient saturates, yielding
\begin{equation}\label{extreme-k}
\alpha = \frac{14}{30\pi} \lambda_P \quad{\rm for} \quad k^{-1} \ll \lambda_P \ll L.
\end{equation}
This result is derived from (\ref{high-k}) by expanding the integrand in its righthand side in powers of 
$\lambda_P$ around $\lambda_P=0$.  
Remarkably, it differs from the long distance behavior reported in Ref.~\cite{EPL}, 
which was derived from the analysis of corrugation for perfectly reflecting plates~\cite{Emig}.
In fact, the high-$k$ expression (\ref{high-k}) holds when the roughness length scale 
$1/k$ is much smaller than both $\lambda_P$ and $L$. A different result is obtained 
when $\lambda_P$ rather than $1/k$ is the shortest length scale. In order to see it,
we derive from our general result an expression for $G$ valid at this limit: 
\begin{align}\label{G}
G(k)= &-\frac{\hbar A}{8\pi^2}\,\frac{1}{L^5\,q}\int_0^{\infty}\frac{dK e^{-2K}}{1-e^{-2K}}
\int_{|K-q|}^{K+q}dK'\\
\times  & 
\frac{(KK')^2+\frac{1}{4}(K^2+K'{}^2-q^2)^2}
{1-e^{-2K'}} \quad {\rm for} \; \lambda_P \to 0.\nonumber
\end{align}
Numerical integration of (\ref{G}) agrees with the expression of Emig {\it et al.}~\cite{Emig} 
for arbitrary values of $q=kL$;  
$K$ has the same meaning already discussed in connection with (\ref{high-k}) while
$K'$ corresponds to the longitudinal component of the imaginary wavevector 
associated with the diffracted wave.  

\begin{figure}[ptb]
\centerline{\psfig{figure=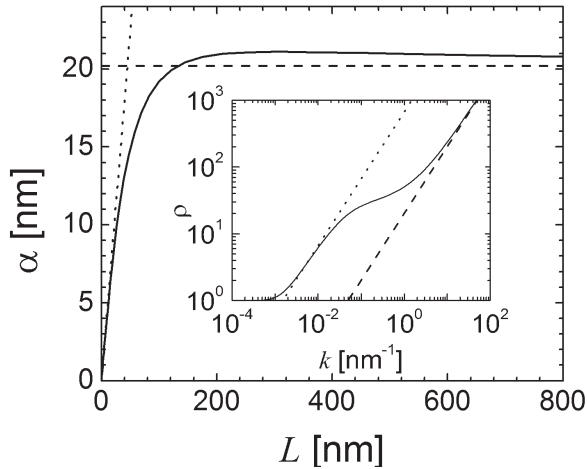,width=9cm}}\caption{Variation of the 
angular coefficient 
$\alpha$ versus $L$ for $\lambda_P=136$nm. The analytical result for 
$k^{-1} \ll L \ll \lambda_P$ is shown as the dotted line and for
$k^{-1} \ll \lambda_P \ll L$ as the dashed line.   
A comparison between this second result (dashed straight line) 
and the exact $\rho(k)$ (solid line) is shown in the inlet for $L=2 \mu$m. 
The analytical result $\rho=L k/3$ predicted by the model of perfect reflectors
(dotted line) is valid only in the intermediate range $ \lambda_P\ll k^{-1} \ll L$.}
\end{figure}

In order to discuss the regime $\lambda_P\ll 1/k \ll L$, we now take the high-$k$ limit 
of the right-hand side of (\ref{G}).
Due to the presence of the exponential factor $\exp(-2K)$, 
the dominant contribution comes from the corner 
$K\stackrel{\scriptstyle <}{\scriptstyle\sim}1,$ $K'\sim q$  
of the rectangle associated to the integration region.  
We may thus neglect $\exp(-2K')$ and 
recover the long distance limit of \cite{EPL}: 
\begin{eqnarray}
\label{perfect}
G(k)&=&-\frac{2}{3\pi^2} \frac{\hbar A q}{L^5} \int_0^{\infty} dK \frac{K^3
\,e^{-2K}}{1-e^{-2K}}=-\frac{\pi^2}{360}\hbar A\frac{q}{L^4}, \nonumber \\
\rho&=&\frac{1}{3}L\,k \quad{\rm for} \quad \lambda_P \ll k^{-1} \ll L. 
\end{eqnarray}
In summary, the long-distance behavior is given by (\ref{extreme-k}) 
when $1/k\ll \lambda_P\ll L,$ and by (\ref{perfect}) when $\lambda_P\ll 1/k\ll L$.
The cross-over between these two regimes is shown in the inlet of Fig.~3, where
we plot $\rho$ as a function of $k$ for $L=2 \mu$m. 
The model with perfect reflectors fails when $1/k\ll \lambda_P$ because
Fourier components with $K\stackrel{\scriptstyle <}{\scriptstyle\sim}1\ll K_p$,
for which the plates behave as perfect reflectors,
are diffracted into components with $K'\sim q\gg K_p$ 
which are no longer perfectly reflected by the plates. 

In order to reach definite conclusions about the accurate evaluation of the 
roughness correction, the first crucial step is to measure the roughness spectrum 
$\sigma({\bf k})$. Once in possession of this experimental input,
one may calculate the roughness correction from the second-order 
perturbation formula ({\ref{main}).
In order to go further in the present letter, we consider the simple model of
a Gaussian roughness spectrum~\cite{Maradudin}
\begin{equation}
\sigma\lbrack\mathbf{k}]=a^{2}\pi\ell_{C}^{2}\exp\left(
-\frac{\mathbf{k}^{2}\ell_{C}^{2}}{4}\right),  
\label{Gaussian}
\end{equation}
$a^2$ is the roughness variance and $\ell_C$ the correlation length. 
Using this model, we now illustrate the preceding results by giving scaling laws
obtained in some limiting cases.

As we increase the distance $L$ between two given plates, characterized 
by the length scales $\lambda_P$ and $\ell_C$, 
the roughness correction decreases according to power laws that may be derived from 
Eqs.~(\ref{main}) and  (\ref{Gaussian}). 
We first consider the case of very smooth surfaces $\lambda_P\ll \ell_C$. 
The correction at short distances $L\ll \lambda_P \ll \ell_C$ may thus be calculated from 
the PFA Eq.~(\ref{eqPFA}) and leads to $\Delta = 3 a^2/L^2$. 
The PFA still holds when  $L$ is increased beyond the plasma wavelength  into
the intermediate range $\lambda_P \ll L \ll \ell_C$ giving $\Delta = 6 a^2/L^2$.
As we increase the distance further, the correction decreases at a slower rate. When 
$\lambda_P \ll \ell_C \ll L$, we find from (\ref{main}), (\ref{perfect}) and (\ref{Gaussian})
the power law $\Delta = 2\sqrt{\pi}\, a^2/(\ell_C L)$, which
represents a correction larger than the PFA result by a factor $L/\ell_C \gg 1$.
The last two results may be derived from the study of perfect reflectors~\cite{EPL,Emig}. 

On the other hand, if we start with a pair of very rough surfaces $\ell_C\ll \lambda_P$, 
we find a completely different behavior as the distance $L$ is increased \cite{comment}.
The short distance limit is still governed by the PFA power law $1/L^2$
but in the intermediate range $\ell_C\ll L \ll \lambda_P$ 
we now find from Eq.~(\ref{kgde-plasmon})
$\Delta = 2.7 \sqrt{\pi}\, a^2/(\ell_C L)$.
For very long distances $\ell_C\ll \lambda_P\ll L$, 
saturation leads to a faster decrease of the correction,
and from Eq.~(\ref{extreme-k}) we find
$\Delta =\left(14/5\sqrt{\pi}\right)
\left(\lambda_P/\ell_C\right)\left(a^2/L^2\right)$. 
This is a $1/L^2$ decrease, as in the PFA, but with an additional factor 
of the order of $\lambda_P/\ell_C \gg 1$. 

In conclusion, we have computed the second-order response function $G(k)$
for arbitrary values of the plasma wavelength $\lambda_P$, distance $L$ and 
roughness wavevector $k.$ This allows for a reliable calculation of the roughness 
correction up to second order in the profiles $h_1$ and $h_2$. 
We have derived analytical results in some limiting cases.
In particular, we have discussed the limits of long distances 
and short roughness wavelengths and shown that their relation to the model of
perfect reflectors is much richer than previously thought~\cite{EPL,Emig}. 
The PFA results are recovered as the limiting case of very long roughness 
wavelengths. 
Our present analysis proves that they systematically underestimate 
the roughness correction \cite{semiclass}. 
This has to be taken into account in the search for bounds 
on new weak forces in the submillimeter range \cite{search_nwf}. 

We thank Cyriaque Genet and Marc-Thierry Jaekel for discussions.
PAMN thanks Instituto do Mil\^enio de 
Informa\c c\~ao Qu\^antica and CNPq for partial 
finantial support.


\begin{thebibliography}{99}
\bibitem{foot} Note that the result of Ref. \cite{Maradudin}
has to be corrected by a global factor 2 in order to pass this test.
This correction was made in Ref. \cite{EPL} which therefore showed the
correct behavior at the short distance limit.
\bibitem{comment}  This calculation relies on the dielectric function 
corresponding to the plasma model. For very rough surfaces it is not 
obvious that this is still a good description of the metallic plates.
\end{thebibliography}
\end{document}